\documentclass[sigconf]{acmart}

\AtBeginDocument{
  }

\renewcommand\footnotetextcopyrightpermission[1]{}

\begin{document}

\title{Human-AI Schema Discovery and Application for Creative Problem Solving}

\author{Sitong Wang}
\email{sw3504@columbia.edu}
\orcid{0000-0002-1401-1371}
\affiliation{
  \institution{Columbia University}
  \city{New York}
  \state{NY}
  \country{USA}
}

\begin{abstract}
Humans often rely on underlying structural patterns—schemas—to create, whether by writing stories, designing software, or composing music.
Schemas help organize ideas and guide exploration, but they are often difficult to discover and apply, especially in complex or unfamiliar domains. 
My Ph.D. research develops a framework for human-AI schema discovery and application to support creative problem solving. 
I design systems that support users in sensemaking over examples to abstract schemas, and in operationalizing schemas into human-AI co-creative workflows for application. 
This research offers insights into how schema-guided interaction can make implicit knowledge more accessible and actionable, advancing more transparent and collaborative human-AI systems.
\end{abstract}

\keywords{schema discovery and application, human-AI interaction, creativity and productivity support tools, agentic AI}

\maketitle
\pagestyle{plain} 

\section{Introduction}
Humans are driven to create—whether through writing stories, designing software, or composing music.
Underlying these creative acts are schemas: structures that capture the essential components of a concept and the relationships among them~\cite{gick1983schema}.
Schemas such as the hero’s journey in storytelling~\cite{campbell2008hero}, design patterns in software development~\cite{gamma1995design}, and chord progressions in music~\cite{fox2013chord} guide creators by embodying key structural principles.
When made explicit, schemas provide reusable frameworks that support purposeful exploration, flexible adaptation, and knowledge transfer across domains.
They help organize complex tasks, scaffold ideation, and make expert strategies accessible to novices—making them powerful tools for creative problem solving.

Despite their importance, schemas can be difficult to discover and apply.
Schema induction—identifying abstract structures from examples—requires recognizing deep, structural similarities often hidden by surface differences.
It is cognitively demanding, requiring abstraction: distilling essential components and relationships from richly varied instances.
For example, uncovering the hero’s journey across Star Wars, The Lord of the Rings, and The Hunger Games is far from obvious—it requires seeing beyond genre, character, and setting to identify shared narrative structures.
Schema application is also challenging—schemas are abstract, applying them involves grounding components in specific contexts, filling in missing details, and adapting relationships to new goals.
Even with a well-known schema like the hero’s journey, creating a story such as Harry Potter still demands extensive ideation, iteration, and adaptation.

My Ph.D. research explores how human-AI schema discovery and application can support creative problem solving.
At the core of this work is supporting sensemaking over examples to extract useful abstractions, and then grounding those abstractions in user contexts through interactive workflows.
In earlier work, I manually derived schemas from expert examples and designed human-AI workflows to help users apply them across a variety of creative tasks.
While this enabled more meaningful exploration, it was often slow and labor-intensive—highlighting the need for intelligent, interactive support.
This laid the foundation for my current work: building interactive systems that facilitate both schema discovery and application through human-AI collaboration.
For schema discovery, the key lies in enabling interactive abstraction—supporting iterative reasoning over examples with AI to surface reusable structure.
For schema application, I focus on helping users instantiate schemas as co-creative workflows with AI: structured loops of divergence and convergence that guide exploration of the design space.
Together, these approaches aim to make schemas more accessible and usable as scaffolding for creative problem solving.

As AI becomes increasingly integrated into complex, open-ended tasks, we must move beyond systems that simply generate outputs.
Instead, AI should support users in exploring problem spaces, learning from expert strategies, and developing creative solutions.
Schemas offer a promising foundation by revealing structural patterns that guide reasoning and adaptation across contexts.
My research contributes a framework for making schemas interactively discoverable and usable, opening new possibilities for more insightful and empowering human-AI collaboration.

\section{Manual Schema Discovery and Human-AI Schema Application}
\begin{figure*}
\includegraphics[width=0.75\textwidth]{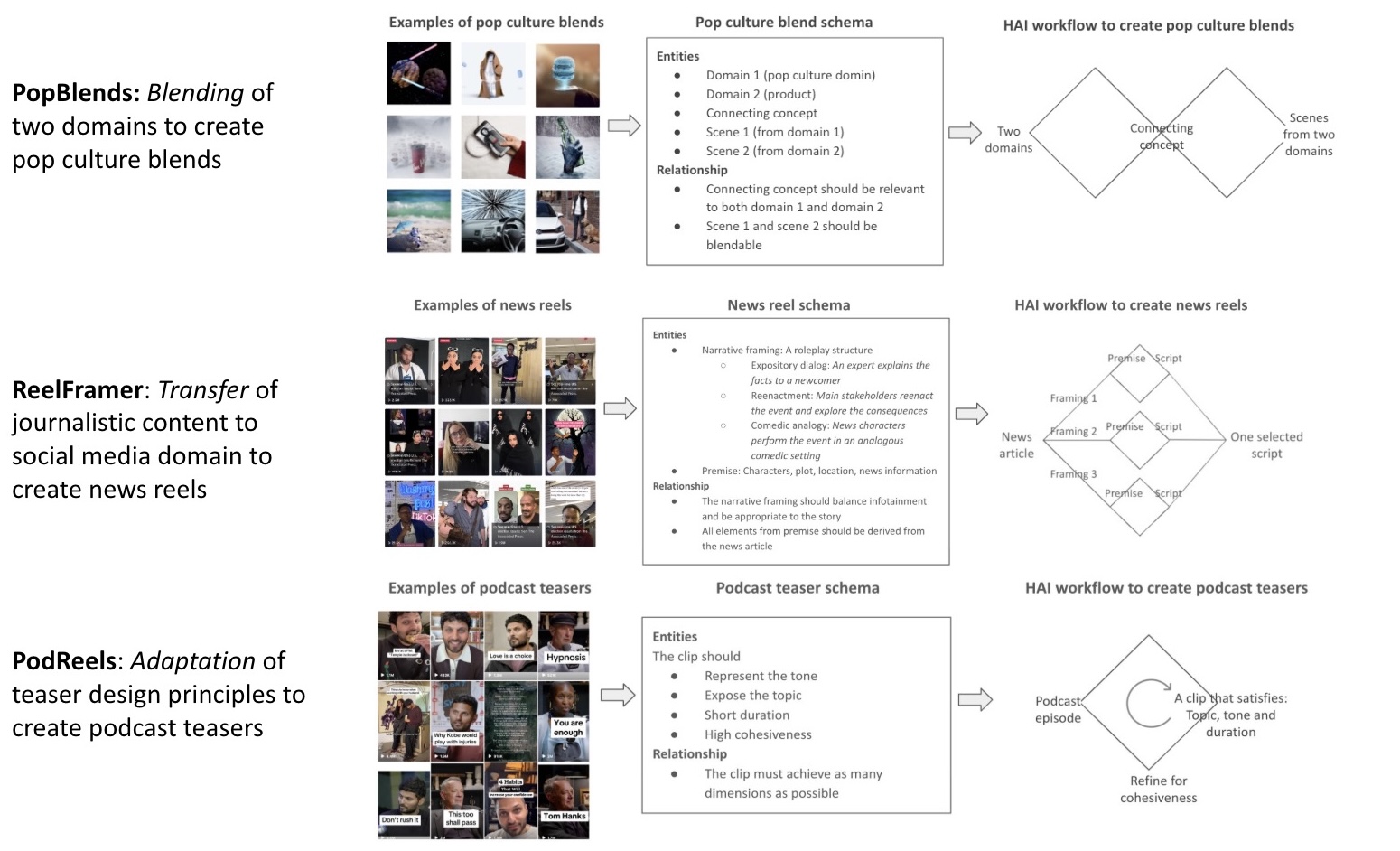}
\caption{Prior Work: Manual schema discovery and human-AI schema application}
\end{figure*}

In my prior work~\cite{popblends,reelframer,podreels}, I manually discovered schemas and built workflows to enable human-AI schema application.
For schema discovery, I primarily identified schemas by analyzing successful examples. 
I also consulted domain experts and referred to relevant theoretical frameworks as complementary methods.
For schema application, I manually translated each schema into a human-AI co-creation workflow. 
This often involved iterative rounds of divergence and convergence to help creators find sets of components that satisfy the relationship constraints.

\textbf{PopBlends~\cite{popblends}: Blending of two domains to create pop culture blends}
Pop culture is powerful for communication.
Social media often features pop culture blends—images that creatively blend a product with a pop culture domain.
To support creating such blends, we began by abstracting schema from examples.
We manually annotated 100 examples, labeling key elements such as the product being promoted, the pop culture reference, the according scenes, and—most importantly—the shared elements connecting the two domains.
Our process was informed by conceptual blending theory, which suggests that effective blends rely on a meaningful connecting concept between domains. 
Building on the schema, we developed PopBlends, an AI-powered workflow that supports creating pop culture blends. 
PopBlends follows a two-stage diverge-converge process to help users ground components and refine their relationships.
Given two domains, PopBlends first identifies potential connecting concepts, then surfaces relevant scenes and images tied to the connecting concept. 
Users can then select and blend scenes from both domains into a final composite.
A user study showed that by leveraging schemas to build conceptual bridges, PopBlends helped creators discover twice as many creative blends with half the mental effort compared to a baseline.

\textbf{ReelFramer~\cite{reelframer}: Transfer of journalistic content to social media domain to create news reels}
News reels are an emerging format for delivering news on social media—short videos created based on existing news articles. 
These reels are often short, narrative-driven, colloquial, and entertaining, differing significantly from traditional news articles in tone and structure. 
However, journalists are not trained to produce such videos, and there is no established design pattern they can follow.
To support journalists in creating news reels, we began by analyzing successful examples to infer the schema. 
We collected 50 examples and conducted a co-analysis with experts in journalism and TikTok. 
From this, we identified two key components: narrative framings (expository dialogue, reenactment, and comedic analogy) and premises (characters, plot, location, and key news information).
The narrative framing must strike a balance between being informative and entertaining, and it should fit the story. 
Meanwhile, the premise should be grounded in the news article.
Based on these insights, we developed a human-AI workflow called ReelFramer, which helps journalists apply the schema in practice. 
Given a news article, ReelFramer guides users through exploring the three narrative framings, expanding each by identifying appropriate premise elements, and iteratively writing scripts until a good one is found.
In our user study, journalists reported that ReelFramer scaffolded the otherwise difficult process of translating news articles into reels by surfacing and operationalizing the narrative structures—ultimately making the process more accessible and enjoyable.

\begin{figure*}
\includegraphics[width=1\textwidth]{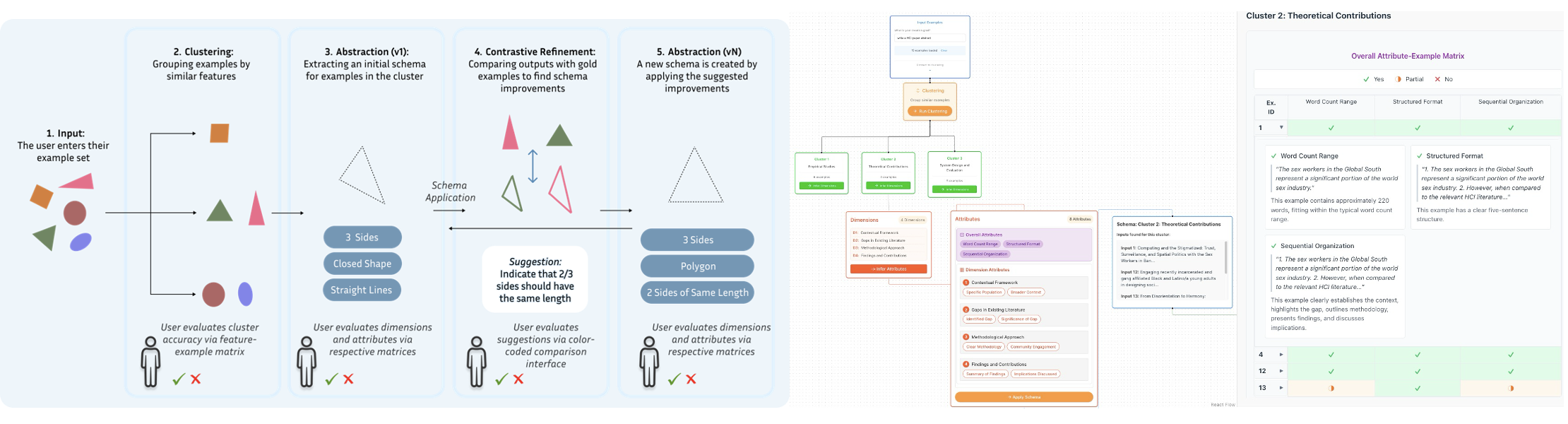}
\caption{Schemex: Facilitating human-AI schema discovery}
\end{figure*}

\textbf{PodReels~\cite{podreels}: Adaptation of teaser design principles to create podcast teasers}
Video podcast teasers are short videos shared on social media to generate interest in full podcast episodes.
They play a crucial role in helping video podcasters reach new audiences and grow their follower base.
However, selecting engaging clips from hour-long episodes requires significant mental effort, making the teaser creation process both time-consuming and cognitively demanding.
We analyzed the successful examples to find the schema: the clip should represent the tone, expose the topic, and have a short duration and high cohesiveness.
An ideal podcast teaser must achieve as many of these dimensions as possible.
We then baked the insights into an interactive workflow to help creators apply this schema: first, it finds the clip that satisfies the dimensions of the desired topic, tone, and duration, and then refines it for cohesiveness.
AI assists in searching for clips, suggesting three clip windows based on multiple user-defined content criteria. 
Creators make the final selection of the clip and also refine and edit the window. 
A user study showed that by facilitating schema application, PodReels enabled podcasters to create teasers in roughly half the time and with significantly reduced cognitive load compared to a baseline.

\section{Facilitating Human-AI Schema Discovery}
Building on my manual experience with schema discovery, I explored how to facilitate human-AI schema discovery to help people abstract patterns from examples using Schemex~\cite{schemex}.

Schema discovery is challenging due to the high degree of ambiguity involved in identifying hidden structures.
Take inducing schemas from the HCI paper abstracts as an example.
First, structural patterns are often obscured by surface-level variation: different examples achieve similar goals through different means. 
For instance, motivation sentences may vary widely in topic, phrasing, and style, yet they all serve the same purpose.
Second, examples within a set may follow different underlying schemas. A system paper and a study paper, for example, contain different components; although related, their schemas are distinct. Trying to derive a single schema for both can lead to overgeneralization.
Third, schemas are difficult to evaluate. Their quality often becomes apparent only when applied to multiple examples to test fit and generalizability.
Lastly, schemas must strike a balance between specificity and generalizability: overly abstract schemas provide insufficient guidance, while overly specific ones fail to transfer across contexts.

To address these challenges, we frame schema induction as an iterative sensemaking process.
Drawing on advances in AI reasoning, we investigate how AI can scaffold the core cognitive steps of clustering, abstraction, and contrastive refinement:
1) Clustering: Starting with a set of examples, AI identifies latent groupings based on structural similarity and organizes examples accordingly.
2) Abstraction: Within each cluster, AI infers underlying dimensions and derives both overall and dimension-specific attributes.
3) Contrastive Refinement: The current schema is used to generate outputs, which are then compared to real (``gold'') examples. These comparisons inform further refinement of the schema.
Throughout this process, humans play a central role as reflective evaluators—they make critical judgments about coherence and utility, direct the exploration, and retain decision-making agency.

To operationalize this approach, we developed Schemex, an interactive visual workflow for schema discovery.
Schemex supports the sensemaking process through AI-assisted reasoning and visualizations that keeps users grounded in real examples.
Users begin by specifying a goal (e.g., “Write HCI paper abstracts”) and providing a set of examples (e.g., CHI best paper abstracts).
These examples can be multimodal; Schemex uses models like GPT-4V and Whisper to convert images or videos into structured textual representations.
The system clusters the examples based on structural similarity and presents them as interactive visual nodes.
Users can explore these clusters to inspect shared features and example-level mappings, validating the coherence of each cluster.
From there, users select a cluster of interest (e.g., empirical studies) and prompt the system to extract key structural dimensions (e.g., Motivation, Method, Findings).
A visualized matrix shows how strongly each example reflects each dimension, supported by inline explanations and citations.
Users can request inferences about dimension-level attributes (e.g., “Research Gap” under Motivation) as well as overall attributes (e.g., tone, length, style).
They assess the coherence of these attributes across examples and examine implementation details through contextual snippets.
Users then apply the derived schema: given a new input (e.g., a paper title), the system generates content guided by the schema.
Schemex enables users to compare generated outputs with original examples through a color-coded dimensional analysis interface.
Differences are highlighted, and the system suggests refinements that users can incorporate.
This iterative cycle—observe, apply, compare, refine—continues until users arrive at a coherent, actionable schema tailored to their goal.

In our user study, participants reported significantly richer and deeper insights, as well as greater confidence in the schemas developed with Schemex, compared to those created using a baseline tool.
They appreciated how Schemex supported sensemaking in an intuitive and interactive manner—allowing them to easily drill into examples to uncover hidden structures and helping them validate schema grounding through example-level visualizations.
\section{Facilitating Human-AI Schema Application}
Given a schema derived from the discovery process, the next step is to help users apply it to their own context through an interactive workflow.
To support this, I propose SchemaBuilder, a tool that transforms schemas into human-AI co-creation workflows.

Applying schemas is challenging because they are typically abstract, while real-world creation and problem solving require concrete details.
To be useful, schemas must be grounded in specific contexts, identifying relevant entities and establishing appropriate relationships among them.
This process often demands a multi-stage workflow to thoroughly explore the design space.
For example, even when guided by a well-known schema like the Hero’s Journey, one must still develop many narrative details before producing a specific story like Harry Potter.

We frame the schema application as a two-stage process:
1) Workflow Specification: Given a schema, the system derives a structured workflow description. This outlines key steps, decision points, and transformations—providing a clear, detailed blueprint for execution.
2) Workflow Execution: This specification is then translated into a functional prototype. Recent advances in coding agents—such as Cline—make it easier to move from specifications to working interactive systems. 
In initial experiments, we found that once a workflow is well-specified, agents like Cline can reliably generate usable interfaces that support real-time interaction.

This shifts the main bottleneck from execution to design: the critical challenge lies in producing high-quality workflow specifications.
To address this, we envision extending the contrastive refinement approach from schema discovery to workflow authoring:
SchemaBuilder takes a schema and example outputs, generates a candidate workflow using AI reasoning, and executes it to produce an artifact.
This artifact is then compared against gold examples to identify gaps, such as missing functionality, misaligned steps, or ambiguous logic.
When a bottleneck is identified, the system triggers a flare-and-focus process: first diverging to explore a range of alternatives, then converging on the most coherent and effective solution.
Through this iterative cycle—generate, execute, compare, refine—SchemaBuilder helps users co-develop reliable, goal-aligned workflows that enable end-to-end schema-driven creation.

\section{Future Directions}

\subsection{Multimodal Schema Discovery and Application}
How can we support creators in identifying and applying visual patterns from examples to new designs?
While Schemex supports multimodal input, it reduces examples to textual descriptions, losing crucial information such as composition and visual hierarchy. I aim to extend Schemex into a fully multimodal system capable of visual reasoning to discover and apply visual schemas. 
This would enable the creation of visual artifacts in domains like posters and covers.

This direction focuses on enabling schema-guided visual creation through structured, interpretable workflows. Rather than prompting a model to generate a complete image, I plan to investigate workflows where an agent—guided by user intent and reference examples—constructs and refines designs using design tools via MCP~\cite{anthropic2024mcp}. The agent would identify reusable patterns, retrieve relevant assets, and adapt layouts dynamically to fit the evolving design context. Users would guide this process by selecting examples, articulating goals, and providing iterative feedback. The aim is to move beyond static templates toward adaptive, structure-aware design workflows that empower creators to explore and apply visual schemas in a flexible and meaningful way.

\subsection{Personalized Workflow Through Schema Discovery of Process History}

While shared schemas and generalized workflows are useful, many creators rely on deeply personal strategies shaped by experience, taste, and tacit knowledge. For instance, a photo editor may follow a distinctive sequence of steps that reflects their individual notion of a “good” image. Rather than prescribing one-fits-all workflows, how might we help users reflect on and formalize their own?

I propose to explore how schema discovery can emerge from users’ process histories. By analyzing real-world editing or design sessions, systems could uncover patterns and routines that inform personalized workflows. These individualized schemas could power custom tools, templates, or intelligent agents that reflect a user’s unique style and practice. Building on SchemaBuilder, which generalizes from example outputs, this approach emphasizes personalized sensemaking and automation. The goal is to support user-centered schema discovery that integrates with existing creative routines and expands one's ability to explore the design space.

\subsection{Co-Agency and Agentic Support for Schema Application}

Schema application, especially in complex, multimodal domains, often involves more than just generating content. It requires selecting relevant context, managing creative assets, and adapting structure to meet evolving goals. I am exploring how AI can support these high-effort activities through intelligent assistance with tasks like context curation and asset manipulation.

Building on my prior work JumpStarter~\cite{jumpstarter}, which helps users curate context for personal projects, I aim to extend agentic support to broader schema application tasks across modalities. Central to this effort is the concept of co-agency—a dynamic sharing of control between humans and AI systems. Effective co-agency involves knowing when the agent should take initiative, offer suggestions, automate low-level tasks, or defer to the user. Understanding and designing for this fluid, context-sensitive agency is key to supporting schema-driven creation that remains aligned with user intent and preserves creative autonomy.

\section{Conclusion}
In this paper, I presented my Ph.D. research on human-AI schema discovery and application for creative problem solving. 
I reviewed my prior work where I manually discovered schemas and developed human-AI workflows to help people apply them. 
I then outlined ongoing and future research aimed at enabling schema discovery and application through human-AI collaboration.
This work focuses on supporting sensemaking over examples to extract useful abstractions and translating those into actionable workflows.
Through this research, I aim to develop frameworks and insights that advance the design of transparent, collaborative human-AI systems.

\bibliographystyle{ACM-Reference-Format}
\bibliography{sample-base}

\end{document}